\documentclass[12pt]{article}
\usepackage{amsmath}
\usepackage{amssymb}
\usepackage{epsfig}
\usepackage{euscript}
\usepackage{fancybox}
\usepackage{color}
\def\mathswitchr#1{\relax\ifmmode{\mathrm{#1}}\else$\mathrm{#1}$\fi}

\newcommand {\pslash}{\hbox{$\not\hbox{\kern-2.3pt $p$}$}}

\usepackage{cite}

\usepackage{epic}
\def\alf1{ {\alpha\over\pi} }
\begin{document}
%\input{feynman}
%=======================================================================
\begin{titlepage}
\begin{flushright}
%{\bf MPI-PhT-2002-08}\\
{\bf BU-HEPP-05-07 }\\
{\bf July, 2006}\\
\end{flushright}
%\vspace{0.05cm}
 
\begin{center}
{\Large Exact Quantum Loop Results in the Theory of General Relativity$^{\dagger}$
}
\end{center}

\vspace{2mm}
\begin{center}
%%  {\bf   S. Jadach$^{a,b}$ and B.F.L. Ward$^{c,d}$}
{\bf   B.F.L. Ward}\\
\vspace{2mm}
%{\em $^a$CERN, Theory Division, CH-1211 Geneva 23, Switzerland,}\\
%{\em $^b$Institute of Nuclear Physics,
%        ul. Kawiory 26a, Krak\'ow, Poland,}
%{\em $^c$Werner-Heisenberg-Institut, Max-Planck-Institut fuer Physik,
%Muenchen, Germany,}\\
%{\em $^a$Werner-Heisenberg-Institut, Max-Planck-Institut fuer Physik,
%Muenchen, Germany,}\\
{\em Department of Physics,\\
 Baylor University, Waco, Texas, 76798-7316, USA}\\
%{\em $^c$SLAC, Stanford University, Stanford, California 94309, USA,}\\
%{\em $^b$Department of Physics and Astronomy,\\
%  The University of Tennessee, Knoxville, Tennessee 37996-1200, USA}\\
%{\em $^c$SLAC, Stanford University, Stanford, California 94309, USA,}\\
\end{center}

\vspace{5mm}
\begin{center}
{\bf   Abstract}
\end{center}
We present a new approach to quantum general relativity based on the
idea of Feynman to treat the graviton in Einstein's theory
as a point particle field subject to quantum fluctuations just as
any such field is in the well-known Standard Model of the electroweak and
strong interactions. We show that, by using resummation techniques
based on the extension of the methods of Yennie, Frautschi and Suura
to Feynman's formulation of Einstein's theory, we get calculable
loop corrections that are even free of UV divergences. One further
by-product of our analysis is that we can apply it to a large class
of interacting field theories, both renormalizable and non-renormalizable,
to render their UV divergences finite as well.
We illustrate our results with applications of some phenomenological
interest.
\vspace{10mm}
\vspace{10mm}
\renewcommand{\baselinestretch}{0.1}
\footnoterule
\noindent
{\footnotesize
\begin{itemize}
\item[${\dagger}$]
Work partly supported
% the Polish Government
%grants KBN 2P30225206 and 2P03B17210, the Maria Sk\l{}odowska-Curie
%Joint Fund II PAA/DOE-97-316, and
by the US Department of Energy grant DE-FG02-05ER41399
and by NATO Grant PST.CLG.980342.
%, and by
%Polish Government grant 5P03B09320.
\end{itemize}
}
%\vspace{0.5cm}
%\begin{flushleft}
%{\bf UTHEP-00-0101}\\
%{\bf Jan, 2000}\\
%\end{flushleft}

\end{titlepage}

%=======================================================================
\def\Kmax{K_{\rm max}}\def\ieps{{i\epsilon}}\def\rQCD{{\rm QCD}}
\renewcommand{\theequation}{\arabic{equation}}
\font\fortssbx=cmssbx10 scaled \magstep2
\renewcommand\thepage{}
%\vfill\eject
\parskip.1truein\parindent=20pt\pagenumbering{arabic}\par
\section{\bf Introduction}\label{intro}\par
The many successes of Einstein's classical theory of
general relativity are well-known~\cite{mtw,sw1}.
Given the outstanding success of the Standard Model(SM)~\cite{sm1,thvt1,qcd1}
point particle
quantum field theory for
the other three known forces, the electromagnetic, weak and strong
interactions, where the non-Abelian loop corrections
predicted by the 't Hooft-Veltman~\cite{thvt1} renormalization
theory for Yang-Mills fields~\cite{ym1} have recently been corroborated
by the precision SM tests~\cite{lewwg} at the CERN LEP Collider,
we have to agree that the union of quantum mechanics
and the classical theory of general relativity is
one key piece of unfinished business
left-over for the 21st century. At this writing,
the only accepted complete treatment of quantum general
relativity, superstring theory~\cite{gsw,jp}, involves
\footnote{ Recently, the loop quantum gravity approach~\cite{lpqg1}
has been advocated by several authors, but it has still unresolved
theoretical issues of principle, unlike the superstring theory.
Like the superstring theory, loop quantum gravity
introduces a fundamental length
, the Planck length, as the smallest distance in the theory. This
is a basic modification of Einstein's theory.}
many hitherto unseen degrees of freedom, some at
masses well-beyond the Planck mass, and this latter property is
understandably a bit unsettling to some. Is it possible that
such degrees of
freedom are anything more than a mathematical artifact?
%The situation is then re-emphasized
%by the outstanding success of the Standard Model (SM) of
%the electroweak~\cite{sm1}
%and strong interactions~\cite{qcd1}, where the non-Abelian loop corrections
%predicted by the 't Hooft-Veltman~\cite{thvt1} renormalization
%theory for Yang-Mills fields~\cite{ym1} has recently been corroborated
%by the precision SM tests~\cite{leewg,nobel} at the CERN LEP Collider,
%for example.
\par

Why can we not apply the 't Hooft-Veltman calculus for
non-Abelian loop corrections to quantum general relativity(QGR)?
After all, the Feynman-Faddeev-Popov ghost field technique, so crucial
to the 't Hooft-Veltman renormalization program, was
invented by Feynman~\cite{rpf1,rpf2}
in his pioneering work on Einstein's theory.
Is it really true, as Einstein suggested, that Bohr's quantum mechanics
is just too incomplete to include general relativity in its domain
of applicability? The superstring theory~\cite{gsw,jp} candidate approach
to quantum general relativity would suggest this as well, as one of its
predictions is that one of the basic results in quantum mechanics, the
Heisenberg uncertainty principle, is in fact modified~\cite{gross}.
Here, we take a different view which we base on the original
work of Feynman~\cite{rpf1,rpf2}. The idea is that the graviton field
should be treated as any other point particle field in the successful SM theory.Just like the famous Higgs field, which has a non-zero vacuum expectation
value about which the {\it physical} Higgs field executes quantum fluctuations,
so too the graviton field, the metric tensor $g_{\mu\nu}$, has a
vacuum expectation value, which we will take following Feynman to be
the Minkowski value $\eta_{\mu\nu}$, about which the {\it physical}
graviton particle executes quantum fluctuations.
When these fluctuations are large, the quantum fluctuations dominate
the metric of space-time and give rise to a regime that has been
called a space-time foam~\cite{wheeler}. We do not discuss this
regime in what follows. When the graviton field fluctuations
are small, we expect to be able to calculate perturbatively in them
using the standard Feynman-Schwinger-Tomonaga methods
if we can find the appropriate {\it representation} of the corresponding
Feynman series. It is in finding the latter representation that
we extend the pioneering ideas of Feynman in our new approach.\par

Our basic objective is to use resummation of large higher order
effects to cure the bad UV behavior of Einstein's theory as formulated
by Feynman. There are essentially two kinds of resummation
algebras that have had some significant amount of success in
the precision theoretical work used in comparing the SM predictions
with the precision LEP data. In the first kind, at each order in the
perturbative expansion, only the terms which are being resummed are retained,
so that what one gets is the exact lowest order term
and the resummation of the large terms
from each order of the loop expansion. While the result is an improvement
over the lowest order term, it is intrinsically an approximate expression.
We call such a resummation an ``approximate resummation''. Examples
are the results in Refs.~\cite{sct,css,js}.
The second type of resummation that has proved useful in precision SM
physics has the property that, while one isolates the terms to be resummed
order by order in perturbation theory, one does not drop the
residual terms in those orders so that one ends up with
an exact expression in which some or all large terms from each order
of perturbation theory are resummed. We call this an ``exact resummation''.
It is an exact re-arrangement of the original Feynman series.
Examples are the theory of Yennie, Frautschi and Suura for QED
in Ref.~\cite{yfs}, its extension to Monte Carlo event generators
in Refs.~\cite{yfs1}, and the QCD and QED$\otimes$QCD
exponentiation in Refs.~\cite{qcdyfs,qced},
which are extensions of the YFS theory to non-Abelian gauge theories.
It is this latter type of resummation which we employ for QGR here;
for, we do not wish to drop any of the effects in theory.
For the record, the results in Refs.~\cite{sct,css,js,yfs1} have played
significant roles in precision tests of SM physics.\par

There are good physical reasons why resummation of the YFS type
properly extended to quantum general relativity may help to tame
the bad UV behavior of the latter theory. Indeed, this at first sight
might even seem counter-intuitive, as the YFS type of resummation resums
large infrared (IR), large distance, effects and the bad UV behavior of quantum general relativity is characteristic of the short distance behavior of the
theory. We make two observations here. First,
in the propagation of a particle between a point $x$
and another point $x'$ in the deep Euclidean regime,
the effective mass squared involved in that propagation is large and
negative, turning the normally attractive Newtonian force for large
positive masses into a large repulsion -- we expect such repulsion to
cause the attendant propagation to be damped severely in the exact solutions
of the theory. While we can not solve the theory exactly, if we can re-arrange
the Feynman series by resumming a dominant part of the large repulsion
we can hope to improve greatly the convergence properties of the
Feynman series. Second, in the Feynman loop integration in 4-momentum space,
there are three regimes in which we may obtain the big logs that
represent dominant behavior: the collinear(CL), infrared (IR) and
ultra-violet (UV) regimes. The CL regime is definitely important but
even in Abelian gauge theory we know that it is difficult to resum
into a simple closed form result with exact residuals. The UV regime
carries the renormalization algebra for the theory and will, after being
tamed, provide us with the relationship between the bare and physical
parameters of the theory. Thus, we do not wish to resum the UV regime.
This leaves us the IR regime, for which we do have a representation
, that of the YFS-type, which is an exact re-arrangement with
closed-form results. We can hope that these resummed 1PI vertices will
result in an improved convergence of the theory. Indeed, in Ref.~\cite{yfs},
it has already been pointed-out that YFS resummation in QED leads to
improved UV behavior for the fermion two-point Green function.
Here, we exploit
this phenomenon applied to quantum general relativity; for, as gravity
couples in the IR regime the same way to all particles, we can
hope that the improvement we find will apply to all particles'
two-point functions.\par

We recall for reference that, as pointed-out in Ref.~\cite{wein1},
there are four basic approaches to the bad UV behavior of
QGR:
\begin{itemize}
\item extended theories of gravitation such as supersymmetric theories (superstrings and supergravity~\cite{supergrav})
and loop quantum gravity; 
\item resummation, a new version of which we discuss presently;
\item composite gravitons; and,
\item asymptotic safety -- fixed point theory, recently pursued with
success in Refs.~\cite{laut,reuter2,litim,perc}.
\end{itemize}
Our approach will allow us to make contact both with the extended
theories and with the phenomenological asymptotic safety approach
results in Refs.~\cite{laut,reuter2,litim,perc}.
Moreover, we note that the recent results in Refs.~\cite{don1,cav,sola}
on the large distance behavior of QGR are not inconsistent with our
approach just as chiral perturbation theory in QCD is not inconsistent
with the application of perturbative QCD to short distance QCD effects.
\par

Ultimately, any approach to QGR has to confront experimental tests
for confirmation. In this paper, we will start this process by addressing
some issues in black hole physics, culminating with an answer to the
fate of the final state of Hawking~\cite{hawk1} radiation
by an originally massive black hole. These 'tests' give us some
confidence that our methods may indeed represent a pure union\footnote{{\it We do not modify Einstein's theory at all.} In this way,
we differ from the currently practiced superstring
theory~\cite{gsw,jp} and loop quantum gravity~\cite{lpqg1} approaches
to the bad ultra-violet behavior of quantum general
relativity. If we are successful, it would be a {\it true} union of the
{\it original} ideas of Bohr and Einstein. We believe this warrants
the further study of our approach in its own right.}
of the ideas of Bohr and Einstein, a union which is not in any contradiction
with any well-established experimental or theoretical result.\par

%Our paper is organized as follows. In the next section, we review
%Feynman's formulation of Einstein's theory in the context of the
%Standard Model~\cite{gsw,qcd} of elementary particles. In Section 3,
%we present the theory of resummed QGR. In Section 4, we discuss
%applications of the new approach to black hole physics issues.
%Section 5 contains the summary of our discussion. The Appendix
%contains some technical details.\par

We shall use resummation
based on the extension to quantum general relativity of the
theory developed by Yennie, Frautschi and Suura (YFS)~\cite{yfs}
originally for QED. In Refs.~\cite{yfs1}, we have extended the
YFS methods to the SM electroweak theory and used these
extended methods to achieve
high precision predictions for SM processes at LEP1 and LEP2, which
have played important roles in the precision SM tests of the
electroweak theory~\cite{lewwg}. Recently~\cite{qcdyfs,qced}, we have made a
preliminary extension of the YFS methods to soft gluon effects
in the QCD sector of the SM, with an eye toward the high energy
processes at the LHC. In this paper, we extend the YFS methods
to treat the bad UV behavior of quantum general relativity.\par

%{\it We do not modify Einstein's theory at all.} In this way,
%we differ from the currently practiced superstring theory~\cite{sprstrg} and loop quantum gravity~\cite{lpg1} approaches to to the bad ultra-violet
%behavior of quantum general
%relativity. If we are successful, it would be a {\it true} union of the
%{\it original} ideas of Bohr and Einstein. We believe this warrants
%the further study of our approach in its own right.\par

%We point out here that, in Ref.~\cite{reuter}, a phenomenological
%realization of Weinberg's asymptotic safety~\cite{wberg1} ansatz
%is achieved for the ultraviolet regime of quantum general relativity
%and that in Refs.~\cite{donog1} an effective Lagrangian approach
%to the infrared limit of quantum general relativity, again following
%ideas of Weinberg~\cite{wberg1}, is achieved using methods familiar
%from the chiral effective Lagrangian approach to low energy QCD.
%We should be able to make contact with these phenomenological
%approaches in the respective limits and we will present our progress
%in so doing in the following.\par

More specifically, in Refs.~\cite{bw1,bw2,bw3,bw4}, we have presented
initial discussions of our new approach. Here we present the
detailed extensions and the complete derivations as needed of the results
in Refs.~\cite{bw1,bw2,bw3,bw4}, as well as several new results.
To make this paper self-contained, we start with the defining
Einstein Lagrangian as formulated by Feynman in Refs~\cite{rpf1,rpf2}
in the context of the Standard Model.
This we do in the next section. In Section 3, we develop and explain
the extension of the resummation theory of Yennie, Frautschi and Suura
to quantum general relativity. In Section 4, we
work out some of the implications of our new approach to quantum
general relativity and make contact with related work in the literature.
Section 5 contains our summary remarks and our outlook.
Technical details are relegated to the Appendices.\par

\section{\bf Einstein's Theory as Formulated by Feynman}\label{defn1}\par

In this section, we formulate Einstein's theory following the
approach of Feynman. This will allow us to set our notation
and conventions and to reveal the true issues one confronts
in quantizing the general theory of relativity.

More precisely, if we denote by $L^{\cal G}_{SM}(x)$ the generally
covariant Standard Model Lagrangian
of the electroweak and strong interactions, then the theory
of the currently known elementary particle interactions has the
point particle field theory Lagrangian
\begin{equation}
%\begin{split}
{\cal L}(x) = -\frac{1}{2\kappa^2}\sqrt{-g} R
            + \sqrt{-g} L^{\cal G}_{SM}(x)
\label{lgwrld1}
\end{equation}
where $R$ is the curvature scalar, $-g$ is the
negative of the determinant of the metric of space-time
$g_{\mu\nu}$, $\kappa=\sqrt{8\pi G_N}\equiv
\sqrt{8\pi/M_{Pl}^2}$, where $G_N$ is Newton's constant,
and the SM Lagrangian density, $L^{\cal G}_{SM}(x)$, which is well-known
( see for example, Ref.~\cite{sm1,thvt1,qcd1,barpass} ) when invariance
under local Poincare symmetry is not required,
is readily obtained
from the familiar SM Lagrangian density as follows:
since $\partial_\mu\phi(x)$ is already generally
covariant for any scalar field $\phi$ and since the only derivatives of the
vector fields in the SM Lagrangian density occur in their
curls, $\partial_\mu A^J_\nu(x)-\partial_\nu A^J_\mu(x)$, which are
also already generally covariant, we only need
to give a rule for making the fermionic terms in
usual SM Lagrangian density generally covariant. For this,
we introduce a differentiable structure with $\{\xi^a(x)\}$ as
locally inertial coordinates and an attendant
vierbein field $e^a_\mu\equiv\partial\xi^a/\partial x^\mu$
with indices that carry
the vector representation for the flat locally inertial space, $a$, and for the
manifold of space-time, $\mu$, with the identification of the space-time
base manifold metric as
$g_{\mu\nu}=e^a_\mu e_{a\nu}$ where the flat locally inertial
space indices are to be
raised and lowered with Minkowski's metric $\eta_{ab}$ as usual.
Associating the usual Dirac gamma
matrices $\{\gamma_a\}$ with the flat locally inertial space at x, we define
base manifold Dirac gamma matrices by $\Gamma_\mu(x)=e^a_\mu(x)\gamma_a$.
Then the spin connection, $\omega_{\mu b}^a=-\frac{1}{2}e^{a\nu}\left(
\partial_\mu e^b_\nu-\partial_\nu e^b_\mu\right)+\frac{1}{2}e^{b\nu}\left(
\partial_\mu e^a_\nu-\partial_\nu e^a_\mu\right)
+\frac{1}{2}e^{a\rho}e^{b\sigma}\left(\partial_\rho e_{c\sigma}-\partial_\sigma e_{c\rho}\right)e^c_\mu$ when there is no torsion, allows us to
identify the generally covariant
Dirac operator for the SM fields by the substitution
${i\!\not\!{\partial}} \rightarrow i\Gamma(x)^\mu\left(\partial_\mu +\frac{1}{2}{\omega_{\mu b}}^a{\Sigma^b}_a\right)$, where we have ${\Sigma^b}_a=\frac{1}{4}\left[\gamma^b,\gamma_a\right]$
everywhere in the SM Lagrangian density. This will generate $L^{\cal G}_{SM}(x)$ from the usual SM Lagrangian density $L_{SM}(x)$ as it is
given in Refs.~\cite{sm1,thvt1,qcd1,barpass}, for example.
The Lagrangian in (\ref{lgwrld1}) will now be treated following
the pioneering work of Feynman~\cite{rpf1,rpf2}.
\par

First, we note that, although the SM Lagrangian is known to contain
many point particle fields, as we are studying the basic interplay
between quantum mechanics and general relativity, for pedagogical
reasons, we focus the the simplest aspect of $L^{\cal G}_{SM}(x)$,
namely that part which involves the massive spinless physical Higgs particle
with only its gravitational interactions -- it will presumably be
observed directly at the LHC in the near future~\cite{ATLAS-CMS}. The major
difficulties in developing a consistent quantum theory of
general relativity are all present in this simplification
of (\ref{lgwrld1}), as has been emphasized by Feynman~\cite{rpf1,rpf2}.
We can return to the treatment of the rest of (\ref{lgwrld1})
elsewhere~\cite{elswh}.\par

In this way we are led to consider here the same theory studied by
Feynman in Refs.~\cite{rpf1,rpf2},
\begin{equation}
\begin{split}
{\cal L}(x) &= -\frac{1}{2\kappa^2} R \sqrt{-g}
            + \frac{1}{2}\left(g^{\mu\nu}\partial_\mu\varphi\partial_\nu\varphi - m_o^2\varphi^2\right)\sqrt{-g}\\
            &= \quad \frac{1}{2}\left\{ h^{\mu\nu,\lambda}\bar h_{\mu\nu,\lambda} - 2\eta^{\mu\mu'}\eta^{\lambda\lambda'}\bar{h}_{\mu_\lambda,\lambda'}\eta^{\sigma\sigma'}\bar{h}_{\mu'\sigma,\sigma'} \right\}\\
            & \qquad + \frac{1}{2}\left\{\varphi_{,\mu}\varphi^{,\mu}-m_o^2\varphi^2 \right\} -\kappa {h}^{\mu\nu}\left[\overline{\varphi_{,\mu}\varphi_{,\nu}}+\frac{1}{2}m_o^2\varphi^2\eta_{\mu\nu}\right]\\
            & \quad - \kappa^2 \left[ \frac{1}{2}h_{\lambda\rho}\bar{h}^{\rho\lambda}\left( \varphi_{,\mu}\varphi^{,\mu} - m_o^2\varphi^2 \right) - 2\eta_{\rho\rho'}h^{\mu\rho}\bar{h}^{\rho'\nu}\varphi_{,\mu}\varphi_{,\nu}\right] + \cdots \\
\end{split}
\label{eq1-1}
\end{equation}
Here,
$\varphi(x)$ is the physical Higgs field as
our representative scalar field for matter,
$\varphi(x)_{,\mu}\equiv \partial_\mu\varphi(x)$,
and $g_{\mu\nu}(x)=\eta_{\mu\nu}+2\kappa h_{\mu\nu}(x)$
where we follow Feynman and expand about Minkowski space
so that $\eta_{\mu\nu}=diag\{1,-1,-1,-1\}$.
Following Feynman, we have introduced the notation
$\bar y_{\mu\nu}\equiv \frac{1}{2}\left(y_{\mu\nu}+y_{\nu\mu}-\eta_{\mu\nu}{y_\rho}^\rho\right)$ for any tensor $y_{\mu\nu}$\footnote{Our conventions for raising and lowering indices in the 
second line of (\ref{eq1-1}) are the same as those
in Ref.~\cite{rpf2}.}.
Thus, $m_o$ is the bare mass of our free Higgs field and we set the small
tentatively observed~\cite{cosm1} value of the cosmological constant
to zero so that our quantum graviton has zero rest mass.
We return to this point, however, when we discuss phenomenology.
%Here, our normalizations are such that $\kappa=\sqrt{8\pi G_N}$
%where $G_N$ is Newton's constant.
The Feynman rules for (\ref{eq1-1}) have been essentially worked out by
Feynman~\cite{rpf1,rpf2}, including the rule for the famous
Feynman-Faddeev-Popov~\cite{rpf1,ffp1} ghost contribution that must be added to
it to achieve a unitary theory with the fixing of the gauge
( we use the gauge of Feynman in Ref.~\cite{rpf1},
$\partial^\mu \bar h_{\nu\mu}=0$ ),
so we do not repeat this
material here. We turn instead directly to the issue
of the effect of quantum loop corrections
in the theory in (\ref{eq1-1}).
\par

\section{\bf Resummation Theory for Quantum General Relativity}\label{resum1}\par

In this section, we develop the resummation theory which we wish to
employ in the context of quantum general relativity. We will follow
the approach of Yennie, Frautschi and Suura in Ref.~\cite{yfs1}.
This choice is made possible by the formulation of Feynman for
Einstein's theory, as the entire theory is a local, point particle
field theory, albeit with an infinite number of interaction vertices.
Perturbation theory methods can be relevant because, to any finite order
in the respective Feynman series, {\em only a finite number of these
interaction terms can contribute}.\par

\begin{figure}
\begin{center}
\epsfig{file=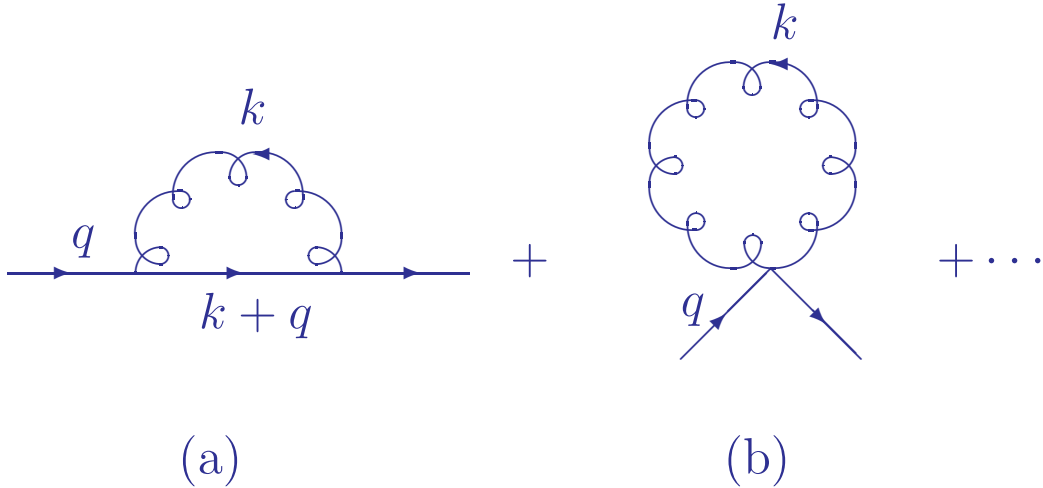,width=140mm}
\end{center}
\caption{\baselineskip=7mm     Graviton loop contributions to the
scalar propagator. $q$ is the 4-momentum of the scalar.}
\label{fig1}
\end{figure}

For the scalar field in (\ref{eq1-1}), consider the
contributions to the 1PI 2-point
function illustrated in Fig.~\ref{fig1}.
We would like
to take advantage of the following physical effect
that is intrinsic in Einstein's formulation of Newton's
law: for large Euclidean momenta, where the squared momentum transfer
in Fig.~\ref{fig1} has a large negative value, the gravitational
self-energy from Newton's law is strongly repulsive, so that
propagation of the particle in this regime should be severely
damped in the {\em exact solutions} of the theory in (\ref{eq1-1}).
This is an intuitive explanation for the success of Weinberg's
asymptotic safety approach as recently realized phenomenologically
in Refs.~\cite{laut,reuter2,litim,perc} and
leads us to try to resum the large parts of
the quantum gravitational loop
corrections in order to improve the convergence of the
respective Feynman series.\par

We, however, do not wish to drop-out pieces of this Feynman series.
We wish to make an {\it exact} re-arrangement of the
series in which some of the large gravitational quantum loop effects
are resummed to all orders in the loop expansion. Which large
gravitational effects shall we resum? In the general Feynman
one-loop integral, enhanced contributions arise from
three regimes:
\begin{itemize}
\item the ultra-violet regime
\item the collinear regime
\item the infrared regime
\end{itemize}
The ultra-violet regime will be treated by the renormalization
program which we seek to establish here. The collinear regime
has been addressed in non-Abelian gauge theories by many authors~\cite{coll}
and we would expect to be able to apply the respective methods to the
improved loop expansion that we seek to establish here as well.
These methods are as yet generally approximate in the sense of our discussion,
they are generically not exact re-arrangements of the Feynman series.
We thus look to the infrared regime, for which exact re-arrangement
of the loop expansion has been achieved by Yennie, Frautschi and Suura (YFS) in
Ref.~\cite{yfs} for Abelian massless gauge theories.
In Refs.~\cite{qcdyfs}, we have shown that the YFS methods can be
extended to non-Abelian gauge theories with the understanding
that only the leading IR singular terms actually exponentiate in the YFS sense
and that the remaining non-leading and genuinely non-Abelian
IR singular terms are treated order by order in the loop expansion.
Physically, resummimg this leading IR singular part of the loop expansion
in quantum general relativity offers the possibility of
improving the convergence of the resummed loop expansion
and curing the long standing problem of the non-renormalizability
of Einstein's theory. That is what we will argue actually
happens in the following.\par

We note here that, already in Ref.~\cite{yfs}, it has been pointed-out that
YFS resummation of the IR effects in QED improves the UV convergence for the Feynman series for QED. This occurs for the electron propagator but not for the
photon propagator and , as the coupling parameter in the soft regime
is just $\alpha_{em}\cong \frac{1}{137}$, the improvement in the convergence
via the electron propagator is very marginal, for it is
the asymptotic behavior $\frac{1}{p^{1+{\cal O}(\alpha_{em})}}$ vs $\frac{1}{p}$ in the deep Euclidean
regime. For quantum general general relativity, we will see below that all
particles' propagators are improved and that the improvement becomes
pronounced in the deep Euclidean regime and causes all propagators
to fall faster than any power of the respective momentum transfer $p$.\par

Returning to Fig.~\ref{fig1}, we write the respective contributions to
the 1PI proper
2-point vertex function, $-i\Sigma(p)$, the proper self-energy contribution
to inverse propagator here, as
\begin{equation}
-i\left(\Delta_F(p)^{-1}-\Sigma(p)\right) = i\sum_{n=o}^{\infty}\Sigma_n(p)
\label{sigma1}
\end{equation}
where $-i\Sigma_n(p)$ is the respective n-loop contribution
with the agreement that for $n=0$ we
have $-\Sigma_0(p)=\Delta_F(p)^{-1}=p^2-m^2
+i\epsilon$
For the latter n-loop contribution, we first represent it as follows:
\begin{equation}
i\Sigma_n(p)= \frac{1}{n!}\int\cdots\int \prod_{i=1}^{n}\frac{dk_i}{k_i^2-\lambda^2+i\epsilon}\rho_n(k_1,\cdots,k_n)
\label{am1}
\end{equation}
where the function $\rho_n$ is symmetric under the interchange
of any two of the n virtual graviton 4-momenta that are exchanged in
(\ref{am1}), by the Bose symmetry obeyed by the spin 2 gravitons
and the symmetry of the respective multiple integration volume.
Here is the point in the discussion where the power of exact rearrangement techniques such as those
in Ref.~\cite{yfs} enters. For the case $n=1$, let $S''_g(k)\rho_0$ represent the leading contribution in the the limit $k\rightarrow 0$ to $\rho_1$.
We have
\begin{equation}
\rho_1(k) = S''_g(k)\rho_0+ \beta_1(k)
\label{indn0}
\end{equation}
where this equation is {\it exact} and serves to define $\beta_1$
if we specify $S''_g(k)$, the soft graviton emission factor, and recall
that 
\begin{equation}
\rho_0=i\Sigma_0(p)=-i\Delta_F(p)^{-1}.
\label{ir-1a}
\end{equation}
This can be determined from the Feynman rules for (\ref{eq1-1}) or
one can also use the off-shell extension of the formulas
in Ref.~\cite{sw1}. We get~\cite{bw1}
\begin{equation}
\begin{split}
S''_g(p,p,k) &= \frac{1}{(2\pi)^4}\frac{i\frac{1}{2}(\eta^{\mu\nu}\eta^{\bar\mu\bar\nu}+
\eta^{\mu\bar\nu}\eta^{\bar\mu\nu}-\eta^{\mu\bar\mu}\eta^{\nu\bar\nu})(-i\kappa p_{\bar\mu})(2ip_\mu)(-i\kappa{p'}_{\bar\nu})(2i{p'}_\nu)}{(k^2-2kp+\Delta+i\epsilon)(k^2-2kp'+\Delta'+i\epsilon)}{\Big|}_{p=p'}\\
           &= \frac{2i\kappa^2p^4}{16\pi^4}\frac{1}{(k^2-2kp+\Delta+i\epsilon)^2}
\end{split}
\label{reasf}
\end{equation}
where $\Delta=p^2-m^2$, $\Delta'={p'}^2-m^2$. To see this, from Fig.~\ref{fig1}, note that the Feynman rules give us the following result
\begin{equation}
\begin{split}
i\Sigma_1(p)&=\Big\{-\frac{\int d^4k}{(2\pi)^4}i v_3(p,p-k)_{\mu\bar{\mu}}\frac{i}{(p-k)^2-m^2+i\epsilon}i v_3(p-k,p')_{\nu\bar{\nu}}\\
&\qquad\qquad \frac{i\frac{1}{2}(\eta^{\mu\nu}\eta^{\bar\mu\bar\nu}+
\eta^{\mu\bar\nu}\eta^{\bar\mu\nu}-\eta^{\mu\bar\mu}\eta^{\nu\bar\nu})}{k^2-\lambda^2+i\epsilon}\\
&\qquad - \frac{\int d^4k}{2(2\pi)^4}i v_4(p,p')_{\mu\bar{\mu};\nu\bar{\nu}}\frac{i\frac{1}{2}(\eta^{\mu\nu}\eta^{\bar\mu\bar\nu} +
\eta^{\mu\bar\nu}\eta^{\bar\mu\nu} - \eta^{\mu\bar\mu}\eta^{\nu\bar\nu})}{k^2-\lambda^2+i\epsilon}\Big\}\Big{|}_{p=p'},
\end{split}
\label{ir-1}
\end{equation}
where we have defined from the Feynman rules the respective 3-point($h\varphi\varphi$ and 4-point($hh\varphi\varphi$) vertices 
\begin{equation}
\begin{split}
i v_3(p,p')_{\nu\bar{\nu}} &=-i\kappa\left(p_\nu{p'}_{\bar\nu}+p_{\bar\nu}{p'}_\nu-g_{\nu\bar{\nu}}(pp'-m^2)\right)\\
i v_4(p,p')_{\mu\bar{\mu};\nu\bar{\nu}} &= -4i\kappa^2[(pp'-m^2)(\eta_{\mu\bar{\nu}}\eta_{\bar{\mu}\nu}+\eta_{\bar{\mu}\bar{\nu}}\eta_{\mu\nu}-\eta_{\mu\bar{\mu}}\eta_{\nu\bar{\nu}}) \\
& \qquad\qquad - (p^{\mu'}{p'}^{\nu'}+p^{\nu'}{p'}^{\mu'})\{\eta_{\mu\mu'}(\eta_{\bar{\mu}\nu}\eta_{\nu'\bar{\nu}}+\eta_{\bar{\mu}\bar{\nu}}\eta_{\nu'\nu} - \eta_{\nu'\bar{\mu}}\eta_{\nu\bar{\nu}})\\
&\qquad\qquad + \eta_{\bar{\mu}\mu'}(\eta_{\mu\nu}\eta_{\nu'\bar{\nu}}+\eta_{\mu\bar{\nu}}\eta_{\nu'\nu} - \eta_{\nu'\mu}\eta_{\nu\bar{\nu}})\}]
\end{split}
\label{ir-2}
\end{equation}
using the standard conventions so that p is incoming and p' is outgoing for 
the scalar particle momenta at the respective vertices. In this way, we 
see that we may isolate the IR dominant part of $i\Sigma_1(p)$ by 
the separation
\begin{equation}
\begin{split}
\frac{1}{k^2-2kp+\Delta+i\epsilon}&=-\frac{\Delta}{(k^2-2kp+\Delta+i\epsilon)^2}+\frac{1}{k^2-2kp+i\epsilon}\\
&\qquad -\frac{2\Delta^2}{(k^2-2kp+\Delta+i\epsilon)^2(k^2-2kp+i\epsilon)}\\
&\qquad -\frac{\Delta^3}{(k^2-2kp+\Delta+i\epsilon)^2(k^2-2kp+i\epsilon)^2}\\
&\qquad +\sum_{n=2}^\infty(-1)^n\frac{\Delta^n}{(k^2-2kp+i\epsilon)^{n+1}}
\end{split}
\label{ir-3a}
\end{equation}
from which we can see that the first term on the RHS gives, upon insertion into (\ref{ir-1}), the IR-divergent
contribution for the coefficient of the lowest order inverse propagator
for the on-shell limit $\Delta\rightarrow 0$. The second term does not produce an IR-divergence and the remaining terms vanish faster than $\Delta$ in the on-shell limit so that they do not contribute to the field renormalization factor which we seek to isolate. In this way we get finally
\begin{equation}
\begin{split}
i\Sigma_1(p) &=
\Big\{-\frac{\int d^4k}{(2\pi)^4}(-2i\kappa p_\mu p_{\bar\mu}+i\delta{v_3(p,p-k)_{\mu\bar{\mu}}})(\frac{-i\Delta}{(k^2-2kp+\Delta+i\epsilon)^2}+i~R\Delta_F(k,p))\\
&\qquad\qquad (-2i\kappa {p}'_\nu {p'}_{\bar\nu}+i\delta{v_3(p'-k,p')_{\nu\bar{\nu}}})\frac{i\frac{1}{2}(\eta^{\mu\nu}\eta^{\bar\mu\bar\nu}+
\eta^{\mu\bar\nu}\eta^{\bar\mu\nu}-\eta^{\mu\bar\mu}\eta^{\nu\bar\nu})}{k^2-\lambda^2+i\epsilon}\\
&\qquad - \frac{\int d^4k}{2(2\pi)^4}i v_4(p,p')_{\mu\bar{\mu};\nu\bar{\nu}}\frac{i\frac{1}{2}(\eta^{\mu\nu}\eta^{\bar\mu\bar\nu} +
\eta^{\mu\bar\nu}\eta^{\bar\mu\nu} - \eta^{\mu\bar\mu}\eta^{\nu\bar\nu})}{k^2-\lambda^2+i\epsilon}\Big\}\Big{|}_{p=p'}\\
&\qquad = \Big\{\frac{\int d^4k}{(2\pi)^4}[(-i\kappa p_{\bar\mu})(2ip_\mu)\frac{-i\Delta}{(k^2-2kp+\Delta+i\epsilon)^2}(-i\kappa {p'}_{\bar\nu})(2i{p'}_\nu)\\
&\qquad\qquad \frac{i\frac{1}{2}(\eta^{\mu\nu}\eta^{\bar\mu\bar\nu}+
\eta^{\mu\bar\nu}\eta^{\bar\mu\nu}-\eta^{\mu\bar\mu}\eta^{\nu\bar\nu})}{k^2-\lambda^2+i\epsilon}+\frac{(2\pi)^4\beta_1(k)}{k^2-\lambda^2+i\epsilon}]\Big\}\Big{|}_{p=p'},
\end{split}
\label{ir-3b}
\end{equation}
which agrees with (\ref{indn0},\ref{ir-1a},\ref{reasf}) with
\begin{equation}
\begin{split}
R\Delta_F(k,p)&=\frac{1}{k^2-2kp+i\epsilon}
-\frac{2\Delta^2}{(k^2-2kp+\Delta+i\epsilon)^2(k^2-2kp+i\epsilon)}\\
&\qquad -\frac{\Delta^3}{(k^2-2kp+\Delta+i\epsilon)^2(k^2-2kp+i\epsilon)^2}\\
&\qquad +\sum_{n=2}^\infty(-1)^n\frac{\Delta^n}{(k^2-2kp+i\epsilon)^{n+1}},\\
i\delta{v_3(p,p-k)_{\mu\bar{\mu}}}&=i v_3(p,p-k)_{\mu\bar{\mu}}-\{-2i\kappa p_\mu p_{\bar\mu}\},\\
\beta_1(k)&= \Big\{-\frac{1}{(2\pi)^4}(-2i\kappa p_\mu p_{\bar\mu}+i\delta{v_3(p,p-k)_{\mu\bar{\mu}}})[\frac{-i\Delta}{(k^2-2kp+\Delta+i\epsilon)^2}\\
&\qquad\qquad +i~R\Delta_F(k,p)]
(i\delta{v_3(p'-k,p')_{\nu\bar{\nu}}})\{i\frac{1}{2}(\eta^{\mu\nu}\eta^{\bar\mu\bar\nu}+
\eta^{\mu\bar\nu}\eta^{\bar\mu\nu}-\eta^{\mu\bar\mu}\eta^{\nu\bar\nu})\}\\
&\qquad\qquad -\frac{1}{(2\pi)^4}(-2i\kappa p_\mu p_{\bar\mu}+i\delta{v_3(p,p-k)_{\mu\bar{\mu}}})(i~R\Delta_F(k,p))\\
&\qquad\qquad (-2i\kappa {p}'_\nu {p'}_{\bar\nu})\{i\frac{1}{2}(\eta^{\mu\nu}\eta^{\bar\mu\bar\nu}+
\eta^{\mu\bar\nu}\eta^{\bar\mu\nu}-\eta^{\mu\bar\mu}\eta^{\nu\bar\nu})\}\\
&\qquad\qquad -\frac{1}{(2\pi)^4}(i\delta{v_3(p,p-k)_{\mu\bar{\mu}}})(\frac{-i\Delta}{(k^2-2kp+\Delta+i\epsilon)^2})\\
&\qquad\qquad (-2i\kappa {p}'_\nu {p'}_{\bar\nu})\{i\frac{1}{2}(\eta^{\mu\nu}\eta^{\bar\mu\bar\nu}+
\eta^{\mu\bar\nu}\eta^{\bar\mu\nu}-\eta^{\mu\bar\mu}\eta^{\nu\bar\nu})\}\\
&\qquad\qquad - \frac{1}{2(2\pi)^4}i v_4(p,p')_{\mu\bar{\mu};\nu\bar{\nu}}\{i\frac{1}{2}(\eta^{\mu\nu}\eta^{\bar\mu\bar\nu} +
\eta^{\mu\bar\nu}\eta^{\bar\mu\nu} - \eta^{\mu\bar\mu}\eta^{\nu\bar\nu})\}\Big\}\Big{|}_{p=p'}.
\end{split}
\label{ir-4a}
\end{equation}
\par
One can see that the
result in (\ref{reasf}) differs from the corresponding result in QED in
eq.(5.13) of Ref.~\cite{yfs} by the replacement of the electron charges
$e$ by the gravity charges $\kappa p_{\bar\mu},~\kappa{p'}_{\bar\nu}$
with the corresponding replacement of the photon propagator numerator
$-i\eta_{\mu\nu}$ by the graviton propagator numerator $i\frac{1}{2}(\eta^{\mu\nu}\eta^{\bar\mu\bar\nu}+
\eta^{\mu\bar\nu}\eta^{\bar\mu\nu}-\eta^{\mu\bar\mu}\eta^{\nu\bar\nu})$.
That the squared modulus of these gravity charges grows quadratically
in the deep Euclidean regime is what makes their effect therein in the quantum
theory of general relativity fundamentally different from the effect
of the QED charges in the deep Euclidean regime of QED, where the latter
charges are constants order-by-order in perturbation theory.\par

Indeed, proceeding recursively, we write
\begin{equation}
\rho_{n}(k_1,\cdots,k_n)=S''_g(k_n)\rho_{n-1}(k_1,\cdots,k_{n-1})+\beta^{(1)}_n(k_1,\cdots,k_{n-1};k_n)
\label{indn1}
\end{equation}
where here the notation indicates that the residual $\beta^{(1)}_n$ does not
contain the leading infrared contribution for $k_n$ that is given by the
first term on the RHS of (\ref{indn1})\footnote{ We stress that it may contain
in general other IR singular contributions.}. We iterate (\ref{indn1}) to get
\begin{equation}
\begin{split}
\rho_{n}(k_1,\cdots,k_n)&=S''_g(k_n)S''_g(k_{n-1})\rho_{n-2}(k_1,\cdots,k_{n-2})\\
&+S''_g(k_n)\beta^{(1)}_{n-1}(k_1,\cdots,k_{n-2};k_{n-1})\\
&+S''_g(k_{n-1})\beta^{(1)}_{n-1}(k_1,\cdots,k_{n-2};k_n)\\
&+\{-S''_g(k_{n-1})\beta^{(1)}_{n-1}(k_1,\cdots,k_{n-2};k_n)+\beta^{(1)}_n(k_1,\cdots,k_{n-1};k_n)\}
\end{split}
\label{indn2}
\end{equation}
The symmetry of $\rho_n$ implies that the quantity in curly brackets is
also symmetric in the interchange of $k_{n-1}$ and $k_n$. We indicate this
explicitly with the notation
\begin{equation}
\{-S''_g(k_{n-1})\beta^{(1)}_{n-1}(k_1,\cdots,k_{n-2};k_n)+\beta^{(1)}_n(k_1,\cdots,k_{n-1};k_n)\} = \beta^{(2)}_n(k_1,\cdots,k_{n-2};k_{n-1},k_n).
\label{indn3}
\end{equation}
\par

Repeated application of (\ref{indn1}) and use of the symmetry of
$\rho_n$ leads us finally to the {\it exact} result
\begin{equation}
\begin{split}
\rho_n(k_1,\cdots,k_n)&=S''_g(k_1)\cdots S''_g(k_n)\beta_0\\
&+\sum_{i=1}^{n}S''_g(k_1)\cdots S''_g(k_{i-1})S''_g(k_{i+1})\cdots S''_g(k_n)\beta_1(k_i) \\
&+ \cdots + \sum_{i=1}^{n}S''_g(k_i)\beta_{n-1}(k_1,\cdots,k_{i-1},k_{i+1},\cdots,k_n)+\beta_n(k_1,\cdots,k_n)
\end{split}
\label{indn3a}
\end{equation}
where the case n=1 has already been considered in (\ref{indn0})
with $\rho_0\equiv \beta_0$. Here, we defined as well $\beta^{(i)}_i \equiv \beta_i$.

We can use the symmetry of the residuals $\beta_i$ to re-write $\rho_n$
as
\begin{equation}
\rho_n(k_1,\cdots,k_n)=\sum_{perm}\sum_{r=0}^{n}\frac{1}{r!(n-r)!}\prod_{i=1}^rS''_g(k_i)\beta_{n-r}(k_{r+1},\cdots,k_n)
\label{indn4}
\end{equation}
so that we finally obtain, upon substitution into (\ref{am1}),
\begin{equation}
i\Sigma_n(p)=\sum_{r=0}^{n}\frac{1}{r!(n-r)!}\left(\int\frac{d^4k~S''_g(k)}{k^2-\lambda^2+i\epsilon}\right)^r\int\prod_{i=1}^{n-r}\frac{d^4k_i}{{k_i}^2-\lambda^2+i\epsilon}\beta_{n-r}(k_1,\cdots,k_{n-r}).
\label{indn5}
\end{equation}
With the definition
\begin{equation}
-B''_g(p)= \int\frac{d^4k~S''_g(k)}{k^2-\lambda^2+i\epsilon}
\label{indn6}
\end{equation}
and the identification
\begin{equation}
i\Sigma'_r(p) =\frac{1}{r!}\int\prod_{i=1}^{r}\frac{d^4k_i}{k_i^2-\lambda^2+i\epsilon}\beta_r(k_1,\cdots,k_r)
\label{indn7}
\end{equation}
we introduce the result (\ref{indn5}) into (\ref{sigma1}) to get
\begin{equation}
\begin{split}
-i\left(\Delta_F(p)^{-1}-\Sigma(p)\right) &= i\sum_{n=0}^{\infty}\sum_{r=0}^n\Sigma'_{n-r}(p)\frac{(-B''_g(p))^r}{r!}\\
&= ie^{-B''_g(p)}\sum_{\ell=0}^{\infty}\Sigma'_\ell(p)\\
&= -ie^{-B''_g(p)}\left(\Delta_F(p)^{-1}-\sum_{\ell=1}^{\infty}\Sigma'_\ell(p)\right).
\end{split}
\label{indn8}
\end{equation}
In this way, our resummed {\it exact} result for the complete propagator
in quantum general relativity is seen to be~\cite{bw1,bw2,bw3}
\begin{equation}
i\Delta'_F(p)|_{\text{resummed}} = \frac{ie^{B''_g(p)}}{(p^2-m^2-\Sigma'_s(p)+i\epsilon)}
\label{indn9}
\end{equation}
where
\begin{equation}
\Sigma'_s(p)\equiv \sum_{\ell=1}^{\infty}\Sigma'_\ell(p).
\label{indn10}
\end{equation}
\par

Some observations are in order before we turn to the consequences of (\ref{indn9}). First, we have not modified Einstein's theory at all. This means
we are developing a very conservative approach to treat the UV
behavior of of quantum general relativity. This makes our approach
interesting in its own right, as we have noted in the Introduction.
Second, because we did not modify the theory, what we have done
is necessarily gauge invariant, as the original theory was gauge invariant.
Third, the IR-improved $\Sigma'_s(p)$ is already organized in a loop expansion by our derivation of (\ref{indn10}). We expect therefore to be able
to treat it perturbatively when the physics allows us to so do.\par

To see the effect of the exponential factor in (\ref{indn9}), we evaluate
the exponent as follows for Euclidean momenta ( see Appendix 1 for
the details of the attendant evaluation~\footnote{See also Ref.~\cite{yfs},
where this result can be inferred from its eq.(5.17) by the substitution
$e^2 \rightarrow -\kappa^2 p^2$ therein, as we have indicated above,
where $p\equiv k$ here.})
\begin{equation}
B''_g(k) = \frac{\kappa^2|k^2|}{8\pi^2}\ln\left(\frac{m^2}{m^2+|k^2|}\right).
\label{expnt}
\end{equation}
The latter result establishes the advertised behavior: in the deep
Euclidean regime, the resummed propagator falls faster than any finite
power of $|k^2|$. This is exactly the type of behavior we need to
tame the bad UV behavior of quantum general relativity.\par

We have in fact shown in Ref.~\cite{bw1} that the exponentially
damped behavior in the the propagator in (\ref{indn9}), which holds
for all particles because gravity couples in the infrared universally
to all particles, leads to the UV finiteness of quantum general
relativity, which is completely consistent with asymptotic safety~\cite{wein1}.
The proof is given explicitly in Ref.~\cite{bw1} -- see especially pages
6-8 of the latter reference -- for completeness, we record it in
Appendix 3 here.
In the next section, we turn to some of the further consequences
of the improved propagator behavior and UV finiteness we have found in
our new approach to quantum general relativity.\par

\section{Resummed Quantum Gravity and Newton's Law: Some Consequences}

An immediate consequence
of our new UV finite quantum loop results for QGR is
that we can make exact, UV finite, predictions for the
quantum loop corrections~\cite{bw1,bw2,bw3,bw4} to Newton's law.
These results are
then unique because we do not modify Einstein's theory or
quantum mechanics to obtain them and we have no
free parameters. We now present our prediction for the
quantum loop corrections to Newton's law in this Section.
\par

\begin{figure}
\begin{center}
\epsfig{file=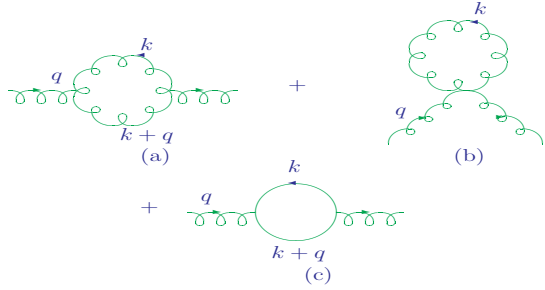,width=77mm,height=38mm}
\end{center}
\caption{\baselineskip=7mm  The graviton((a),(b)) and its ghost((c)) one-loop contributions to the graviton propagator. $q$ is the 4-momentum of the graviton.}
\label{fig2}
\end{figure}

\begin{figure}
\begin{center}
\epsfig{file=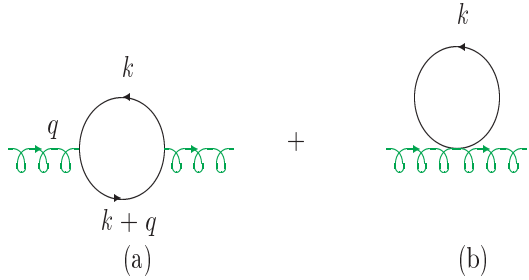,width=77mm,height=38mm}
\end{center}
\caption{\baselineskip=7mm  The scalar one-loop contribution to the
graviton propagator. $q$ is the 4-momentum of the graviton.}
\label{fig3}
\end{figure}

Specifically, consider the diagrams in Figs.~\ref{fig2} and \ref{fig3}.
These graphs have a superficial degree of divergence
in the UV of +4 and in the usual treatment
of the theory they are well-known to generate
a UV divergence in the respective 1PI 2-point function
for the coefficient of $q^4$, a divergence that thus can not be
removed by the standard field and mass renormalizations.
Any successful treatment of the UV behavior of QGR must therefore render this
divergence finite. Indeed,
when the graphs Figs. 2 and 3 are computed
in our resummed quantum gravity
theory, this is precisely what happens.\par

For example, consider the graph in Fig.~3a. When we use our resummed
propagators, we get
(here, $k\rightarrow (ik^0,\vec k)$ by Wick rotation, and we work in the
transverse-traceless space){\small
\begin{equation}
\begin{split}
i\Sigma(q)^{1a}_{\bar\mu\bar\nu;\mu\nu}&=i\kappa^2\frac{\int d^4k}{2(2\pi)^4}
\frac{\left(k'_{\bar\mu}k_{\bar\nu}+k'_{\bar\nu}k_{\bar\mu}\right)e^{\frac{\kappa^2|{k'}^2|}{8\pi^2}\ln\left(\frac{m^2}{m^2+|{k'}^2|}\right)}}
{\left({k'}^2-m^2+i\epsilon\right)}\\
   &\qquad   \frac{
\left(k'_{\mu}k_{\nu}+k'_{\nu}k_{\mu}\right)e^{\frac{\kappa^2|k^2|}{8\pi^2}\ln\left(\frac{m^2}{m^2+|k^2|}\right)}}
{\left(k^2-m^2+i\epsilon\right)}.
\end{split}
\label{eq2p}
\end{equation}}\noindent
%This integral is convergent! So is that for Fig 3b.
%\begin{equation}
%\label{prop1}
%\end{equation}
We see explicitly that the exponential damping in the deep
Euclidean regime has rendered the graph in Fig.~3a finite in the UV.
Similarly are all the graphs in Figs. 2 and 3 UV finite when we use
our respective resummed propagators to compute them.\par

To evaluate the effect of the corrections in Figs.~\ref{fig2} and \ref{fig3}
on the graviton propagator, we continue to work in the
transverse, traceless space and isolate the effects from
Figs.~\ref{fig2} and \ref{fig3} on the coefficient of the
$q^4$ in the graviton propagator denominator,
\begin{equation}
q^2 +\frac{1}{2}q^4\Sigma^{T(2)}+i\epsilon,
\label{prop}
\end{equation}
so that we need to evaluate the transverse, traceless self-energy
function $\Sigma^T(q^2)$ that
follows from eq.(\ref{eq2p}) for Fig. 3a and its analogs for Figs. 3b
and 2 by the standard methods. Here, we work in the
expectation that, in consequence to the newly UV finite calculated
quantum loop effects in Figs. 2 and 3, the Fourier
transform of the graviton propagator that enters Newton's law,
our ultimate goal here, will receive support from
%poles $q^2 =0$ as usual
%and
%from $q^2=bM_{Pl}^2$ with $|b|<<1$. We will therefore work
from $|q|^2<<M_{Pl}^2$. We will therefore work
in the limit that $q^2/M_{Pl}^2$ is relatively small,~$\lesssim .1$,
for example. This will allow us to see the dominant effects
of our new finite quantum loop effects. In other words, we will
work to $\sim 10\%$ accuracy in what follows.\par
 
First let us dispense with the contributions from Figs. 2b
and Fig. 3b. These are independent of $q^2$ so that
we use a mass counter-term to remove them and set the graviton mass
to $0$. %\footnote{ 
Following the suggestion of Feynman in Ref.~\cite{rpf2}, 
we will change this to a small non-zero value below
to take into account the recently established small value of the
cosmological constant~\cite{cosm1}. See also the discussion in 
Ref.~\cite{novello} where it is 
shown that the quantum fluctuations in the exact de Sitter metric implied by 
the non-zero cosmological constant correspond in general to a 
mass for the graviton. 
Here, as we expand about a flat background, we take this effect into account 
as a small infrared regulator for the graviton. The deviations from flat space 
in the deep Euclidean region that we study due to the observed value
of the cosmological constant are at the level of $e^{10^{-61}}-1$! This is
safely well beyond the accuracy of our methods.\par
%}.\par

Returning to Fig.~3a, when we project onto the transverse, traceless
space, that is to say, the graviton helicity space $\{e^{\mu\nu}(\pm2)=\epsilon^\mu_\pm\epsilon^\nu_{\pm},\, \text{where}\, \epsilon^\nu_{\pm}=\pm(\hat{x}\pm i\hat{y})/\sqrt{2}\, \text{when}\, \\ \hat{x},~\hat{y}\, \text{are purely space-like and} (\vec{\hat{x}},\vec{\hat{y}},\vec{q}/|\vec{q}|)\, \text{form a right-handed coordinate
basis}\}$,  we get (see the Appendix 2) the result
\begin{equation}
i\Sigma^T(q^2)_{3a}=\frac{-i\kappa^2m^4}{96\pi^2}\int^1_0d\alpha\int^{\infty}_0 dx\frac{x^3(2(x+1)\bar{d}+\bar{d}^2)}{(x+1)^2(x+1+\bar{d})^2}(1+x)^{-\lambda_c x}
\label{eq3p}
\end{equation}
where $\lambda_c=\frac{2m^2}{\pi M_{Pl}^2}$,~$\bar{d}=\alpha(1-\alpha)\vec{q}^2/m^2$ so that
we have made the substitution $x=k^2$ and imposed the mass counter-term
as we noted. We have taken for definiteness $q=(0,\vec{q})$. We also
use $q=|\vec{q}|$ when there is no chance for confusion.
We are evaluating (\ref{eq3p}) in the deep UV where $m^2/q^2<< 1$
and where $q^2/M_{Pl}^2\lesssim 0.1$. In this case,
we get
\begin{equation}
i\Sigma^T(q^2)_{3a}=\frac{-i\kappa^2}{96\pi^2}\left(\frac{|\vec{q}|^2m^2c_1}{3}+\frac{|\vec{q}|^4c_2}{30}\right)
\end{equation}
where
\begin{equation}
\begin{split}
c_1=I_1(\lambda_c)&=\int^{\infty}_0dx x^3(1+x)^{-3-\lambda_c x}\\
c_2=I_2(\lambda_c)&=\int^{\infty}_0dx x^3(1+x)^{-4-\lambda_c x}.
\end{split}
\end{equation}
Using the usual field renormalization, we see that
Fig. 3a makes the contribution
\begin{equation}
i\tilde\Sigma^T(q^2)_{3a}\cong \frac{-i\kappa^2|\vec{q}|^4c_2}{2880\pi^2}
\end{equation}
to the transverse traceless graviton proper self-energy function.\par

Turning now to Figs. 2, the pure gravity loops, we use a contact
between our work and that of Refs.~\cite{tHvelt1}.
In Refs.~\cite{tHvelt1}, the entire set of one-loop divergences
have been computed for the theory in (\ref{eq1-1}).
The basic observation is the following. As we work only to the leading
logarithmic accuracy in $\ln \lambda_c$, it is sufficient to identify
the correspondence between the divergences as calculated in the n-dimensional
regularization scheme in Ref.~\cite{tHvelt1} and as they would occur
when $\lambda_c\rightarrow 0$. This we do by comparing our result
for (\ref{eq3p}) when $q^2\rightarrow 0$ with the corresponding result in
Ref.~\cite{tHvelt1} for the same theory. In this way we see that
we have the correspondence
\begin{equation}
-\ln\lambda_c \leftrightarrow \frac{1}{2-n/2}.
\label{uv1}
\end{equation}
This allows us to read-off the leading log result for the
pure gravity loops directly from the results in Ref.~\cite{tHvelt1}.
Since $-\ln\lambda_c = \ln {M_{Pl}}^2-\ln m^2 -\ln\frac{2}{\pi}$,
we see that our exponentiated propagators have cut-off our UV divergences
at the scale $\sim M_{Pl}$ and the correspondence in (\ref{uv1}) shows the usual relation between the effective UV cut-off scale and the pole in $(2-n/2)$
in dimensional regularization. 
\par
Specifically, the result in Ref.~\cite{tHvelt1}, when interpreted as we
have just explained, is that the pure gravity
loops give a factor of 42 times the scalar loops for the coefficient
$a_2$ above when we work in the regime where $|q^2|$ is relatively
small compared to $M_{Pl}^2$. Here, we also take into account the recent significant evidence for a non-zero cosmological constant~\cite{cosm1}, which can be seen to provide a small non-zero rest mass for the graviton, $m_g\cong 3.1\times 10^{-33}$eV, which serves as an IR regulator for the graviton. This is the value of rest mass in $\lambda_c$ which should be used for pure gravitational
loops. See the Appendix 1 for the derivation of the
corresponding infrared exponents.
\par
We note that, for $\lambda_c=0$, the constant $c_2$ is infinite
and, as we have already imposed both the mass and field renormalization
counter-terms, there would be no physical parameter into which
that infinity could be absorbed: this is just another manifestation
that QGR, without our resummation, is a non-renormalizable theory.\par
\par
Using the universality of the coupling of the graviton when
the momentum transfer scale is relatively small compared to $M_{Pl}$,
we can extend the result for the scalar field above to the remaining
known particles in the Standard Model by counting the number of physical
degrees of freedom for each such particle and replacing the mass of the
scalar with the respective mass of that particle. For a massive fermion
we get a factor of 4 relative to the scalar result with the appropriate
change in the mass parameter from $m$ to $m_f$, the mass of that fermion,
for a massive vector,
we get a factor of 3 relative to the scalar result, with the corresponding
change in the mass from $m$ to $m_V$, the mass of that vector, etc.
In this way, we arrive at the result that the denominator of the
graviton propagator becomes, in the Standard Model,
\begin{equation}
q^2+\Sigma^T(q^2)+i\epsilon\cong q^2-q^4\frac{c_{2,eff}}{360\pi M_{Pl}^2},
\label{dengrvprp}
\end{equation}
where we have defined
\begin{equation}
\begin{split}
c_{2,eff}&=\sum_{\text{SM particles j}}n_jI_2(\lambda_c(j))\\
         &\cong 2.56\times 10^4
\end{split}
\label{c2eff}
\end{equation}
with $I_2$ defined above and with $\lambda_c(j)=\frac{2m_j^2}{\pi M_{Pl}^2}$ and~\cite{bw3}
$n_j$ equal to the number of effective degrees of particle $j$ as
already illustrated. In arriving at (\ref{c2eff}), we take the SM
masses as follows: for the now presumed three massive neutrinos~\cite{neut},
we estimate a mass at $\sim 3$ eV; for
the remaining members
of the known three generations of Dirac fermions
$\{e,\mu,\tau,u,d,s,c,b,t\}$, we use~\cite{pdg2002,pdg2004}
$m_e\cong 0.51$ MeV, $m_\mu \cong 0.106$ GeV, $m_\tau \cong 1.78$ GeV,
$m_u \cong 5.1$ MeV, $m_d \cong 8.9$ MeV, $m_s \cong 0.17$ GeV,
$m_c \cong 1.3$ GeV, $m_b \cong 4.5$ GeV and $m_t \cong 174$ GeV and for
the massive vector bosons $W^{\pm},~Z$ we use the masses
$M_W\cong 80.4$ GeV,~$M_Z\cong 91.19$ GeV, respectively.
We note that (see the Appendix 1) when the
rest mass of particle $j$ is zero,
the value of $m_j$ turns-out to be
$\sqrt{2}$ times the gravitational infrared cut-off
mass~\cite{cosm1}, which is $m_g\cong 3.1\times 10^{-33}$eV.
We further note that, from the
exact one-loop analysis of Ref.\cite{tHvelt1}, it also follows
that the value of $n_j$ for the graviton and its attendant ghost is $42$.
For $\lambda_c\rightarrow 0$, we have found the approximate representation
\begin{equation}
I_2(\lambda_c)\cong \ln\frac{1}{\lambda_c}-\ln\ln\frac{1}{\lambda_c}-\frac{\ln\ln\frac{1}{\lambda_c}}{\ln\frac{1}{\lambda_c}-\ln\ln\frac{1}{\lambda_c}}-\frac{11}{6}.
\end{equation}
\par
If we use the standard Fourier transform of the respective graviton
propagator we obtain the improved Newton potential
\begin{equation}
\Phi_{N}(r)= -\frac{G_N M}{r}(1-e^{-ar}),
\label{newtnrn}
\end{equation}
where with
%\begin{equation}
%c_{2,eff} \cong 2.56\times 10^{4}
%\end{equation}
%and, from eq.(8) in Ref.~\cite{bw3},
\begin{equation}
a \cong (\frac{360\pi M_{Pl}^2}{c_{2,eff}})^{\frac{1}{2}}
\end{equation}
we have that
\begin{equation}
a \cong  0.210 M_{Pl}.
\end{equation}
\par
We note that the implied behavior of the running Newton constant, $G_N(k)$,
that corresponds to our resummed graviton propagator denominator,
\begin{equation}
G_N(k)=\frac{G_N}{1+k^2/a^2},
\end{equation}
agrees with the large (Euclidean) $k^2$ limit of $G_N(k)$ 
found by the authors in
Ref.~\cite{reuter2} using the asymptotic safety approach as realized by 
phenomenological exact renormalization group methods -- we agree as well 
on the generic size of $a$. The connection between k and position 
space in our analysis
is given by the usual Fourier transformation method whereas that 
in Ref.~\cite{reuter2} involves a phenomenological parameter which 
is ideally determined self-consistently. Thus, as we will see below, 
while our results and the results
in Ref.~\cite{reuter2} agree on $G_N(k)$ for large values of $k$, our 
forms for the corresponding Newton potential differ in position space: 
we expect our result to hold in the deep Euclidean regime whereas at 
larger distances the result in Ref.~\cite{reuter2} should be preferred.\par
We also note that the behavior of the graviton propagator found by our 
analysis and by that in Ref.~\cite{reuter2} agrees for 
large Euclidean $k^2$ with that
in the $R^2-$quantum gravity theory~\cite{r2gravity}. However, 
unlike the latter theory, in our work  
unitarity has not been lost, as we quantize the theory using the methods
of Refs.~\cite{rpf1,rpf2,thvt1}.\par 
We discuss now two consequences of the improved Newton potential:\\
\subsection{Elementary Particles and Black Holes}
One of the issues that confronts the theory of point particle
fields is that fact that a massive point particle of rest mass $m$ has its
mass entirely inside of its Schwarzschild radius $r_S=2m/M_{Pl}^2$
so that classically it should be a black hole. We expect this
conclusion to be modified by quantum mechanics, where the mass of such
a particle seems readily accessible in experiments.
Note that we distinguish here the uncertainty in the position
of the particle, which is connected to its Compton wavelength when
the particle is at rest, from the accessibility of the mass of that particle,
which is connected to its black hole character.
The situation can be addressed by focusing on the lapse function
in the metric class
\begin{equation}
ds^2 = f(r)dt^2-f(r)^{-1}dr^2 - r^2d\Omega^2,
\label{mclass}
\end{equation}
with
\begin{equation}
f(r)=1-\frac{2G(r)m}{r}
\label{lapse}
\end{equation}
and $G(r)$, using (\ref{newtnrn}), given by
\begin{equation}
G(r)=G_N(1-e^{-ar}).
\label{lapse2}
\end{equation}
We see that the Standard Model massive particles all have the
property that $f(r)$ remains positive as $r$ passes through their
respective Schwarzschild radii and goes to $r=0$, so that the
particle is no longer~\cite{bw2,bw3} a black hole as it was classically.
Refs.~\cite{reuter2,maart} have also found that sub-Planck mass black holes
do not exist in quantum field theory.
\par
\subsection{Final State of Hawking Radiation -- Planck Scale Cosmic Rays}
The situation that then naturally comes to mind is the evaporation of
massive black holes. In Ref.~\cite{reuter2}, 
following Weinberg's~\cite{wein1}
asymptotic safety approach as realized by phenomenological
exact renormalization group methods, it has been shown that the
attendant
running of Newton's constant\footnote{See Ref.~\cite{odint} for a discussion of the gauge invariance issues here.} leads to the lapse function
representation, in the metric class in (\ref{mclass})
\begin{equation}
f(r)=1-\frac{2G(r)M}{r}
\label{massive}
\end{equation}
where $M$ is the mass of the black hole and now
\begin{equation}
G(r)\equiv G_{BR}(r)=\frac{G_Nr^3}{r^3+\tilde{\omega}G_N\left[r+\gamma G_N M\right]}
\label{rnG}
\end{equation}
where $\gamma$ is a phenomenological
parameter~\cite{reuter2} satisfying $0\le\gamma\le\frac{9}{2}$ and
$\tilde{\omega}=\frac{118}{15\pi}$. It can be shown that
(\ref{rnG}) leads as well to the conclusion that black holes with
mass less than a critical mass $M_{cr}\sim M_{Pl}$
have no horizon, as we have argued for massive SM elementary particles.
When we join our result in (\ref{lapse}) onto that in (\ref{rnG})
at the outermost solution, $r_>$, of the equation
\begin{equation}
G_{BR}(r)=G_N(1-e^{-ar}),
\label{match}
\end{equation}
we have a result for the final state of the Hawking process
for an originally very massive black hole: for $r<r_>$, in the
lapse function we
use our result in (\ref{lapse}) for $G(r)$ and for $r>r_>$ we
use $G_{BR}(r)$ for $G(r)$ after the originally massive black hole
has Hawking radiated down to the appropriate scale. For example,
for the self-consistent
value $\gamma=0$ and $0.2=\Omega\equiv\frac{\tilde\omega}{G_NM^2}=\frac{\tilde\omega M_{Pl}^2}{M^2}$ for definiteness we find~\cite{bw4} that the
inner horizon found in Ref.~\cite{reuter2}
moves to negative values of $r$ and
that the outer horizon moves to $r=0$, so that
the entire mass of the originally very massive black hole radiates away
until a Planck scale remnant of mass $M_{cr}'=2.38~M_{Pl}$ is
left~\cite{reuter2,rizzo}
%\footnote{Ref.~\cite{maart} argues as well that the loop
%quantum gravity approach implies that black holes below a critical
%mass do not form, in agreement with Ref.~\cite{reuter2}. Here, we show that the attendant remnants are actually accessible to our universe.}
, which then is completely accessible to our universe. It would be expected
to decay into n-body final states, $n=2,3,\cdots$, leading in general
to Planck scale cosmic rays~\cite{bw3,bw4}. The data in Ref.~\cite{cosmicray,westerhoff}
are not inconsistent with this conclusion, which also agrees with
recent results by Hawking~\cite{hawk2}.\par
\section{Conclusions}
In this paper we have introduced a new paradigm in the history of
point particle field theory: a UV finite theory of the quantum general
relativity. It appears to be a solution to most of the outstanding problems
in the union of the ideas of Bohr and Einstein.
More importantly, it shows that quantum mechanics,
while not necessarily the ultimate theory, is not an incomplete
theory.
\par
Our paradigm does not contradict any known experimental
or theoretical fact; rather, it allows us to better understand
the known physics and, hopefully, to make new testable predictions.
Our paradigm does not contradict string theory or loop quantum gravity,
to the best of our knowledge. In principle, all three approaches to
quantum general relativity should agree in the appropriate regimes,
where we would stress that, unlike what is suggested by the other
two approaches, sub-Planck
scale phenomena do exist in our approach.
Further work on establishing the precise relationship
between the three approaches is in progress.
\par
Evidently, formulations for supergravity theories
in Refs.~\cite{supergrav,vnNh} which were abandoned as complete
theories of quantum gravity because they proved to be non-renormalizable are
now, with the resummation methods of this paper, rendered UV finite and
thus are again phenomenologically interesting in their own
right rather than as low energy approximations to surperstring theory.
Of course, they may still have other problems.
We will pursue this line of phenomenology elsewhere.\par

\section*{Acknowledgements}

We thank Profs. S. Bethke and L. Stodolsky for the support and kind
hospitality of the MPI, Munich, while a part of this work was
completed. We thank Prof. S. Jadach for useful discussions.

\section*{Appendix 1: Evaluation of Gravitational Infrared Exponent}

In the text, we use several limits of the gravitational infrared
exponent $B''_g$ defined in (\ref{indn6}). Here, we present
these evaluations for completeness.\par

We have to consider
\begin{equation}
\begin{split}
-B''_g(p)&= \int\frac{d^4k~S''_g(k)}{k^2-\lambda^2+i\epsilon}\\
         &= \int\frac{d^4k}{(2\pi)^4(k^2-\lambda^2+i\epsilon)}\frac{i\frac{1}{2}(\eta^{\mu\nu}\eta^{\bar\mu\bar\nu}+
\eta^{\mu\bar\nu}\eta^{\bar\mu\nu}-\eta^{\mu\bar\mu}\eta^{\nu\bar\nu})(-i\kappa p_{\bar\mu})(2ip_\mu)(-i\kappa{p'}_{\bar\nu})(2i{p'}_\nu)}{(k^2-2kp+\Delta+i\epsilon)(k^2-2kp'+\Delta'+i\epsilon)}{\Big|}_{p=p'}\\
           &= \frac{2i\kappa^2p^4}{16\pi^4}\int\frac{d^4k}{(k^2-\lambda^2+i\epsilon)}\frac{1}{(k^2-2kp+\Delta+i\epsilon)^2}
\end{split}
\label{gexponent}
\end{equation}
where $\Delta=p^2-m^2$. The integral on the RHS of (\ref{gexponent}) is given by
\begin{equation}
\begin{split}
I&=\int\frac{d^4k}{(k^2-\lambda^2+i\epsilon)}\frac{1}{(k^2-2kp+\Delta+i\epsilon)^2}\nonumber\\
&= \frac{-i\pi^2}{p^2}\frac{1}{x_+-x_-}{\big[}x_+\ln(1-1/(\sqrt{2}x_+))-x_-\ln(1-1/(\sqrt{2}x_-)){\big]}
\end{split}
\end{equation}
with
\begin{equation}
x_\pm=\frac{1}{2\sqrt{2}}\left(\bar\Delta+\bar\lambda^2\pm((\bar\Delta+\bar\lambda^2)^2-4(\bar\lambda^2-i\bar\epsilon))^{\small 1/2}\right)
\end{equation}
for $\bar\Delta=1-m^2/p^2,~\bar\lambda^2=\lambda^2/p^2 \text{and}~\bar\epsilon=\epsilon/p^2$. In this way, we arrive at the results, for $p^2 <0$,
\begin{equation}
B''_g(p)
=\begin{cases}
&\frac{\kappa^2|p^2|}{8\pi^2}\ln\left(\frac{m^2}{m^2+|p^2|}\right),~~m\ne 0\\
&\frac{\kappa^2|p^2|}{8\pi^2}\ln\left(\frac{m_g^2}{m_g^2+|p^2|}\right),~~m=m_g=\lambda\\
&\frac{2\kappa^2|p^2|}{8\pi^2}\ln\left(\frac{m_g^2}{|p^2|}\right),~~m=0,~m_g=\lambda
\end{cases}
\end{equation}
where we have made more explicit the presence of the
observed small mass, $m_g$, of the graviton.\par

\section*{Appendix 2: Evaluation of Gravitationally Regulated Loop Integrals}
In this section we present the derivation of the representations
which we have used in the text in evaluating the gravitationally
regulated loop integrals in Figs.~\ref{fig2},\ref{fig3}.\par
Considering the integrals in Fig.~\ref{fig3} to show the methods, we need the
result for
\begin{equation}
\begin{split}
{\cal I}_{\bar\mu\bar\nu;\mu\nu}&=i\frac{\int d^4k}{(2\pi)^4}
\frac{\left(k'_{\bar\mu}k_{\bar\nu}+k'_{\bar\nu}k_{\bar\mu}\right)e^{\frac{\kappa^2|{k'}^2|}{8\pi^2}\ln\left(\frac{m^2}{m^2+|{k'}^2|}\right)}}
{\left({k'}^2-m^2+i\epsilon\right)}\\
   &\qquad   \frac{
\left(k'_{\mu}k_{\nu}+k'_{\nu}k_{\mu}\right)e^{\frac{\kappa^2|k^2|}{8\pi^2}\ln\left(\frac{m^2}{m^2+|k^2|}\right)}}
{\left(k^2-m^2+i\epsilon\right)}.
\end{split}
\label{a21}
\end{equation}
In the limit that $|q^2|<<M_{Pl}^2$, standard symmetric integration methods
give us, for the transverse parts,
\begin{equation}
{\cal I}_{\bar\mu\bar\nu;\mu\nu}=\frac{i\pi^2}{12}\{g_{\bar\mu\bar\nu}g_{\mu\nu}+permutations\}I_0
\end{equation}
where we have
\begin{equation}
I_0\cong\frac{\int_0^1d\alpha\int_0^\infty dkk^3}{(2\pi)^4}\frac{k^4e^{\lambda_c(k^2/m^2)\ln(m^2/(m^2+k^2))}}{[k^2+m^2+|q^2|\alpha(1-\alpha)]^2}
\end{equation}
and where we used the symmetrization, valid under the respective integral sign,
\begin{equation}
k_{\bar\mu}k_{\bar\nu}k_\mu k_\nu \rightarrow \frac{k^4}{24}\{g_{\bar\mu\bar\nu}g_{\mu\nu}+permutations\}
\label{sym}
\end{equation}
and $\lambda_c=2m^2/(\pi M_{Pl}^2)$.
The integral $I_0$, with the use of the mass counter-term, then leads
us to evaluate the difference,
\begin{equation}
\Delta I=I_0(q)-I_0(0)\cong \frac{\int_0^1 d\alpha \int_0^\infty dx}{2(2\pi)^4}\frac{x^3(x+1)^{-\lambda_c x}}{(x+1)^2(x+1+\bar d)^2}\left(-2\bar d(x+1)-\bar{d}^2\right)
\end{equation}
where we define here $\bar d=|q^2|\alpha(1-\alpha)$.
It is seen that the dominant part of the integrals comes
from the regime where $x\sim 1/(\rho\lambda_c)$ with $\rho= -\ln\lambda_c$,
so that we may finally write
\begin{equation}
\begin{split}
\Delta I&=I_0(q)-I_0(0)\\
&\cong \frac{\int_0^1 d\alpha \int_0^\infty dx}{2(2\pi)^4}\frac{x^3(x+1)^{-\lambda x}}{(x+1)^2(x+1+\bar d)^2}\left(-2\bar d(x+1)-\bar{d}^2\right)\\
&\cong -\frac{|q|^2I_1}{6(2\pi)^4}-\frac{|q|^4I_2}{60(2\pi)^4}
\end{split}
\label{a22}
\end{equation}
where we have defined
\begin{equation}
\begin{split}
I_1(\lambda_c)&=\int^{\infty}_0dx x^3(1+x)^{-3-\lambda_c x},\nonumber\\
I_2(\lambda_c)&=\int^{\infty}_0dx x^3(1+x)^{-4-\lambda_c x}.
\end{split}
\end{equation}
The result (\ref{a22}) has been used in the text.
\par
For the limit in practice, where we have $\lambda_c\rightarrow 0$, we can get
accurate estimates for the integrals $I_1,I_2$ as follows.
Consider first $I_2$. Write $x^3=(x+1-1)^3= (x+1)^3-3(x+1)^2+3(x+1)-1$
to get
\begin{equation}
\begin{split}
I_2(\lambda_c)&=\int^{\infty}_0dx \left((1+x)^{-1}-3(x+1)^{-2}+3(x+1)^{-3}-(x+1)^{-4}\right)(1+x)^{-\lambda_c x}\nonumber\\
&\cong \int^{\infty}_0dx(x+1)^{-1-\lambda_c x} -\frac{11}{6}.
\end{split}
\end{equation}
Use then the change of variable $r=\lambda_c x$ to get, for $\rho=\ln(1/\lambda_c)$,
\begin{equation}
\begin{split}
 \int^{\infty}_0dx(x+1)^{-1-\lambda_c x}&= \int^{\infty}_0dr\frac{e^{-r\ln(r+\lambda_c)-\rho r}}{r+\lambda_c}\\
&=-\ln\lambda_c+\int^{\infty}_0dr\ln(r+\lambda_c)(\ln(r+\lambda_c)+r/(r+\lambda_c)+\rho)e^{-r\ln(r+\lambda_c)-\rho r}\\
&\cong \rho + \int^{\infty}_0dr\sum_{j=0}^\infty\frac{1}{j!}((\rho +1)(\partial/\partial\alpha)^{j+1}+ (\partial/\partial\alpha)^{j+2})(\partial/\partial\rho)^j r^\alpha e^{-\rho r}|_{\alpha=0}\\
&=\rho +\sum_{j=0}^\infty\frac{1}{j!}((\rho +1)(\partial/\partial\alpha)^{j+1}+ (\partial/\partial\alpha)^{j+2})(\partial/\partial\rho)^j\Gamma(\alpha+1)\rho^{-\alpha-1}|_{\alpha=0}\\
&\cong \rho +\frac{-(\rho+1)\ln\rho +\ln^2\rho}{\rho-\ln\rho}\\
&= \rho -\ln\rho -\frac{\ln\rho}{\rho-\ln\rho}.
\end{split}
\end{equation}
This gives us the approximation
\begin{equation}
I_2(\lambda_c)= \rho -\ln\rho -\frac{\ln\rho}{\rho-\ln\rho} - \frac{11}{6}
\end{equation}
%with $\rho=-\ln\lambda_c$
when $\lambda_c\rightarrow 0$, as we noted
in the text.
\par
The integral $I_1$ is a field renormalization constant so, in the usual
renormalization program, we do not need it for most of the applications.
Here, we will discuss it as well for completeness. We get
\begin{equation}
\begin{split}
I_1(\lambda_c)&= \int_0^\infty dx(1+x)^{-\lambda_c x}-3\left(I_2(\lambda_c)+\frac{11}{6}\right)+\frac{5}{2}\nonumber\\
&= \int_0^\infty dx(1+x)^{-\lambda_c x}-3I_2(\lambda_c)-3,
\end{split}
\end{equation}
where, as above, we use
\begin{equation}
\begin{split}
\int_0^\infty dx(1+x)^{-\lambda_c x}&=\frac{\int_0^\infty dr}{\lambda_c}e^{-r\ln(r+\lambda_c)-r\rho}\nonumber\\
&\cong \frac{\int_0^\infty dr}{\lambda_c}\sum_{j=0}^\infty\frac{1}{j!}(\partial/\partial\rho)^j(\partial/\partial\alpha)^jr^\alpha e^{-\rho r}|_{\alpha=0}\nonumber\\
&= \frac{1}{\lambda_c} \sum_{j=0}^\infty\frac{1}{j!}(\partial/\partial\rho)^j(\partial/\partial\alpha)^j\Gamma(1+\alpha)\rho^{-\alpha-1}|_{\alpha=0}\nonumber\\
&\cong  \frac{1}{\lambda_c}\frac{1}{\rho-\ln\rho}.
\end{split}
\end{equation}
Thus, we get
\begin{equation}
I_1(\lambda_c)\cong \frac{1}{\lambda_c}\frac{1}{\rho-\ln\rho} - 3I_2(\lambda_c)-3.
\label{iapprx1}
\end{equation}
\par
Finally, let us show why we can neglect the terms $\bar d$ that were in the denominators of $I_j,~j=1,2$. It is enough to look into the differences
\begin{equation}
\Delta I_j=\frac{\int_0^\infty dx x^3}{(x+1)^j}\left(\frac{1}{(x+1)^2}-\frac{1}{(x+1+\bar d)^2}\right)(x+1)^{-\lambda_c x},~j=1,2
\label{diffints}
\end{equation}
where we note that the integral $I_1$ is absorbed by the standard field renormalization where here for convenience we do this at $|q^2|=0$ when we neglect
$\bar d$ in the denominator of $I_1$ or at the zero of the respective
graviton propagator away from the origin otherwise.
From this perspective, the main integral to examine to illustrate
the level of our approximation becomes
\begin{equation}
\begin{split}
\Delta I_{2}&=\frac{\int_0^\infty dx }{(x+1)^2}\{\frac{(x+1)^{-\lambda_c x}}{(x+1)^2}-\frac{(x+1)^{-\lambda_c x}}{(x+1+\bar d)^2}\}\\
&=\frac{\int_0^\infty dr\, e^{-r\ln(r+\lambda_c)-r\rho}}{(r+\lambda_c)^2}\{\frac{1}{(r+\lambda_c)^2}-\frac{1}{(r+\lambda_c+\sigma)^2}\}\\
&\cong \int_0^\infty dr\int_0^\infty d\alpha_1\alpha_1\int_0^\infty d\alpha_2\alpha_2 e^{-r\ln r-r\rho-\alpha_1(r+\lambda_c)-\alpha_2(r+\lambda_c)}\left(1-e^{-\alpha_2\sigma}\right),
\end{split}
\label{idel2int-1}
\end{equation}
where we have defined $\sigma=\lambda_c\bar d$. The approximation
\begin{equation}
\begin{split}
\left(1-e^{-\alpha_2\sigma}\right)&=2e^{-\alpha_2\sigma/2}\sinh(\alpha_2\sigma/2)\\
&\cong \alpha_2\sigma e^{-\alpha_2\sigma/2}
\end{split}
\label{idel2int-2}
\end{equation}
then allows us to get
\begin{equation}
\begin{split}
\Delta I_{2}
&\cong 4\sigma\frac{\partial^2}{\partial\sigma^2}\int_0^\infty dr\, e^{-r\rho}\left(
1-\frac{\lambda_c+\sigma/2}{r+\lambda_c+\sigma/2}\right)\\
&\cong 2+\rho\sigma+2\rho\sigma(1+\frac{1}{4}\rho\sigma)e^{\rho\sigma/2}(C+\ln(\rho\sigma/2)+\sum_{n=1}^{\infty}\frac{(-1)^n(\rho\sigma/2)^n}{n\;n!})\\
\end{split}
\label{idel2int-3}
\end{equation}
which shows that this difference is indeed non-leading log.
The analogous analysis holds for $\Delta I_1$ as well.
\par

\section*{Appendix 3: Proof of UV Finiteness to All Orders in $\kappa$}

For completeness, in this Appendix we review the proof given in Ref.~\cite{bw1}
that the exponentially damped propagators we found in the text render QGR 
UV finite to all orders in $\kappa$.\par
Let us examine the entire theory from (\ref{eq1-1}) to all orders in
$\kappa$: we write it as 
\begin{equation}
{\cal L}(x) = {\cal L}_0(x)+\sum_{n=1}^{\infty}\kappa^n{\cal L}_I^{(n)}(x)
\label{lgn1}
\end{equation}
in an obvious notation in which the first term is the free 
Lagrangian, including the free part of the gauge-fixing and ghost
Lagrangians and the interactions, including the ghost interactions, 
are the terms of ${\cal O}(\kappa^n), n\ge 1$.\par

Each ${\cal L}_I^{(n)}$ is itself a finite sum of terms:
\begin{equation}
{\cal L}_I^{(n)}(x)=\sum_{\ell=1}^{m_n}{\cal L}_{I,\ell}^{(n)}(x).
\label{lgn2}
\end{equation}
${\cal L}_{I,\ell}^{(n)}$ has dimension $d_{n,\ell}$.
Let $d^M_n=\max_\ell\{d_{n,\ell}\}$. As we have at least three fields
at each vertex, the maximum power of momentum at any vertex in
${\cal L}_I^{(n)}$ is $\bar{d}^M_n=\min\{d^M_n-3,2\}$ and is finite
( here, we use the fact that the Riemann tensor is only second order
in derivatives ). We will use this fact shortly. 

First we stress that, in any gauge,
if $P_{\alpha_1\cdots;\alpha'_1\cdots}$ is the respective
propagator polarization sum for a spinning particle, then the spin 
independence of the soft graviton resummation exponential factor
in (\ref{indn9}) yields the respective resummed improved Born
propagator as
\begin{equation}
i{\cal D}^{(0)}_{F\alpha_1\cdots;\alpha'_1\cdots}(k)|_{\text{resummed}}=\frac{iP_{\alpha_1\cdots;\alpha'_1\cdots}e^{B''_g(k)}}{(k^2-m^2+i\epsilon)},
\label{spnprp}
\end{equation}
so that it is also exponentially damped at high energy in the 
deep Euclidean regime (DER). Our improved Born propagators are then
used throughout the respective resummed loop expansion according to
the standard resummation algebra well-tested in the electroweak
theory~\cite{barpass,lep1}. We will use this shortly as well.\par

Now consider any one particle irreducible vertex $\Gamma_N$ with
$[N]\equiv n_1+n_2$ amputated external legs, where we use the notation
$N=(n_1,n_2)$,
when $n_1 (n_2)$ is the respective number of graviton(scalar) external lines.
We always assume we have Wick rotated.
At its zero-loop order, 
there are only tree contributions which are manifestly UV finite.
Consider the first loop (${\cal O}(\kappa^2)$) corrections to  $\Gamma_N$.
There must be at least one improved exponentially damped propagator
in the respective loop contribution and at most two vertices
so that the maximum power of momentum in the numerator of the
loop due to the vertices is $\max\{2\bar{d}^M_1,\bar{d}^M_2\}$  and is finite. 
The exponentially damped propagator then renders the loop integrals finite 
and as there are only a finite number
of them, the entire one-loop (${\cal O}(\kappa^2)$) contribution is finite.\par

As a corollary, if $\Gamma_N$ vanishes in tree approximation, we can
conclude that its first non-trivial contributions at one-loop are
all finite, as in each such loop the exponentially damped 
propagator which must be present
is sufficient to damp the respective finite order polynomial
in loop momentum that occurs from its vertices by our arguments
above into a convergent integral.\par

\begin{figure}
\begin{center}
\epsfig{file=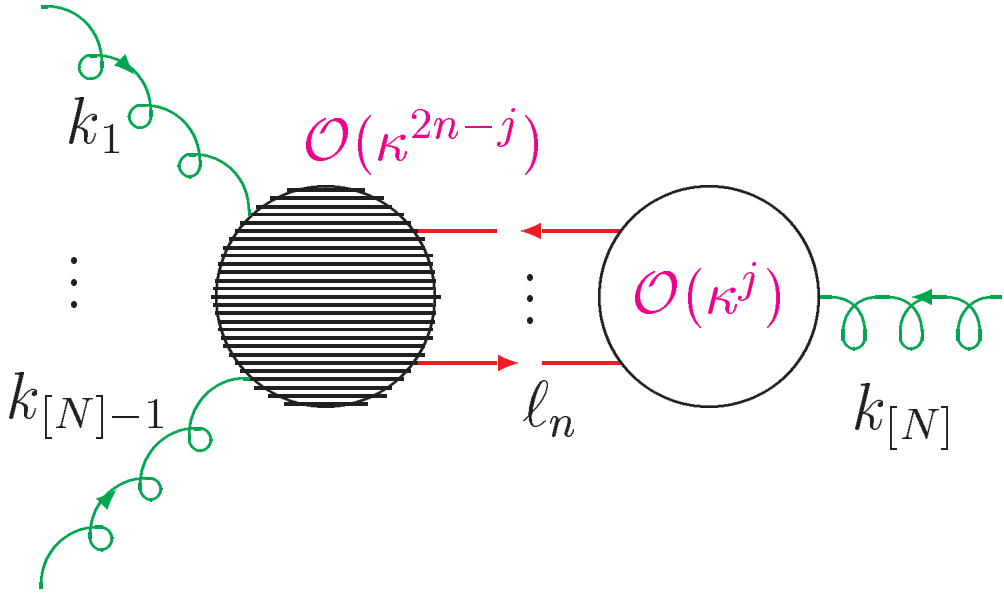,width=140mm}
\end{center}
\caption{\baselineskip=7mm     The typical contribution
we encounter in $\Gamma_N$ at the n-loop level; $\ell_n$ is the n-th
loop momentum and is precisely the momentum of the
indicated resummed improved Born propagator.
}
\label{fig2App}
\end{figure}

As an induction hypothesis suppose all contributions to all $\{\Gamma_N\}$
for m-loop corrections (${\cal O}(\kappa^{2m})), m<n,$ are finite. 
At the n-loop (${\cal O}(\kappa^{2n})$) level, when the exponentially damped
improved Born propagators are taken into account, 
we argue that respective n-loop integrals are finite as follows.
First, by momentum conservation, if $\{\ell_1,\cdots,\ell_n\}$
are the respective Euclidean loop momenta, we may without loss of content
assume that $\ell_n$ is precisely the momentum of one of the
exponentially damped
improved Born propagators. The $n-1$ loop integrations over
the remaining loop variables $\{\ell_1,\cdots,\ell_{n-1}\}$
for fixed $\ell_n$ then produces the contribution of a subgraph
which if it is 1PI is a part of $\Gamma_{N+2}$ and which if it is not 1PI
is a product of the 
contributions to the respective $\{\Gamma_J\}$ and the respective 
improved resummed Born propagator functions.
This is then finite by the induction hypothesis.
Here, $N+2=(n_1+2,n_2)\left((n_1,n_2+2)\right)$ according as the propagator
with momentum $\ell_n$ which we fix as multiplying the remaining subgraph
is a graviton(scalar) propagator, respectively.
The application of standard 
arguments~\cite{Royden} from Lebesgue integration
theory ( specifically, for any two measurable functions $f,g$,
$f\le g$ almost everywhere implies that $\int f \le \int g$ ) 
in conjunction with
Weinberg's theorem~\cite{wein2} guarantees that
this finite result behaves at most as
a finite power of $|\ell_n|$ modulo Weinberg's logarithms 
for $|\ell_n|\rightarrow \infty$. It follows that the remaining
integration over $\ell_n$ is damped into convergence by the
already identified exponentially damped propagator with momentum $\ell_n$.
Thus, each $n$-loop contribution to $\Gamma_N$ is finite,
from which it follows that $\Gamma_N$ is finite at $n$-loop level.
Pictorially, we illustrate the type of situations we have
in Fig.~\ref{fig2App}.
\par

We conclude by induction that all $\{\Gamma_N\}$ in our theory are finite
to all orders in the loop expansion. Of course, the sum of the respective
series in $\kappa$ may very well not actually converge but this issue
is beyond the scope of our work.\par
This completes our Appendix. 
\par

\end{document}